\input{aipcheck}
\documentclass[
    ,final            
  ]
  {aipproc}

\layoutstyle{6x9}

\usepackage{color}

\usepackage[utf8x]{inputenc}
\usepackage{graphicx}
\newcommand*{\mpl}{M_{\rm{Pl}}}
\newcommand*{\lag}{\mathcal{L}}
\newcommand*{\si}{{\rm sign}}

\usepackage{amsmath}

\begin{document}

\title{Probing two approaches to Unified Dark Dynamics}


\author{Jorge L. Cervantes-Cota}{address={Depto. de F\'{\i}sica, Instituto Nacional de Investigaciones
Nucleares, A.P. 18-1027,  11801 M\'{e}xico DF.}}
\author{Alejandro Aviles}{address={Instituto de Ciencias Nucleares, UNAM, M\'exico\\ 
Depto. de F\'isica, Instituto Nacional de Investigaciones Nucleares,  A.P. 18-1027,  11801 M\'{e}xico DF.}}
\author{Josue De-Santiago}{address={Universidad Nacional Aut\'onoma de M\'exico,  04510 M\'exico DF\\ 
Depto. de F\'isica, Instituto Nacional de Investigaciones Nucleares,  A.P. 18-1027,  11801 M\'{e}xico DF.}}


\begin{abstract}
Dark matter and dark energy are essential in the description of the late Universe, since at least the epoch of equality. On the other hand, the inflation is 
also necessary and demands a "dark" component, usually associated to a scalar field that dominated the dynamics and kinematics in the very early Universe.
Yet, these three dark components of standard model of cosmology are independent  from each other, although there are alternative models 
that pursue to achieve a triple unification, or at least a double. In the present work we present an update of two models that we have considered 
in recent years.  The first is the {\it dark fluid} model in which dark matter and dark energy are the same thing, achieving a double unification with specific properties that exactly 
emulate the standard model of cosmology, given the dark degeneracy that exists in the $\Lambda$CDM model.   The second model is given by 
a single $F(X)$ scalar field Lagrangian, with which one is able to model the whole cosmological dynamics, from inflation to today, representing 
a triple unification model.   We highlight the main properties of these models, as well as we test them against known cosmological probes.

\end{abstract}

\maketitle
\section{Introduction}
The standard model of cosmology is based on the existence of dark matter and dark energy, apart from the particle content 
of standard model of particle physics, see ref.  \cite{CervantesCota:2011pn} for a short review. These dark components have been 
dominating the dynamics and kinematics since at 
least the equality epoch, when non-relativistic matter dominated over the relativistic components, and they are essential to 
understand the evolution of both the background and the perturbed cosmos.  However, we have not yet a definitive knowledge 
of their origin nor strong clues on what relationship the dark components  could  have among each other. It is suspected that they 
might share a common origin since the amount of dark matter and dark energy is of the same order of magnitude 
today ($\Omega_{m} \approx \frac{1}{3}\Omega_{de}$), a fact known as coincidence problem \cite{PhysRevD.67.083513}.    

On the other hand, the standard model of cosmology includes an inflationary dynamics at very early times that is important mainly to solve 
the old long-standing puzzles (the horizon and flatness problems and a causal origin of perturbation seeds).  This accelerated dynamics implies 
the existence of  another "dark" (not yet seen) component that is thought to be due to some scalar field dynamics.  The scalar field is then presumed, in (pre-) re-heating,  
to be converted into bosons and fermions that made the Universe material. 

Given the above facts, one identifies three dark (unknown) components of the Universe: dark matter, dark energy,  and the inflationary energy.  Do they have a 
common origin? or least a couple of them? These are questions that have had many answers, as many as a plethora of models in the literature of unified 
dark components, see for instance \cite{Copeland:2006wr,Bertacca:2010ct}.   The task is not simple since we are treating here with very different energy 
scales. By comparing for instance the energy scale of inflation and that of dark energy, that in the limiting case if inflation would have happened at the 
Planck scale, $\rho_{\rm inflation} \sim 10^{122} \rho_{de}$; the most standard energy scale of inflation ($\rho_{\rm inflation} \sim (10^{15} {\rm GeV})^4$) subtracts 
{\it only} 16 orders magnitude to that difference.   On the other hand, 
if reheating took place at the end of inflation, the oscillating scalar field behaves as a dust gas ($p=0$) \cite{Turner:1983he}. The key issue to identify it with dark matter   
is that the reheating process must reduce the field density enough to make it subdominant during the radiation epoch, but not completely, for it to account for the 
right proportion today ($\Omega_{dm} \sim 0.2-0.3$), something that proved to be nontrivial to  achieve in standard reheating schemes \cite{Kofman:1997yn,Liddle:2006qz}.

The above facts indicate some of the difficulties to perform a unification of the different dark components of the Universe. In pursuing it, we 
would like to stress the following simple properties of these components that have led us to propose two different models of unification. Let us start 
mentioning that from the dynamical point of view one needs a dark matter fluid with negligible pressure ($p \ll \rho$, and in fact in the standard 
model $p=0$). As a second property, one requires to have the right proportion of dark matter to baryons, 
$\frac{\Omega_{dm}}{\Omega_{b}} \approx 5$ \cite{Hinshaw:2012fq}. A third property is that the kinematics of dark matter is such that yields potential wells, from 
astrophysical to cosmological scales.  This in turn implies that the effective speed of sound of dark matter is $c_s \ll c$, and in fact in the 
standard model $c_s =0$.   About dark energy we want to remark also three properties: First, it has a particular energy scale, $\rho_{de}$, that dominates the 
background dynamics over all other components since a recent redshift $z \approx 0.5$ \cite{Busca:2012bu}. Second,  it has a pressure proportional to its density 
$p = - \rho_{de}$, and third, it does not seem to cluster in sub-horizon scales.   Finally, the  
inflation dynamics demands a series of tests to be accomplished, such as to yield  a minimum of e-folds of expansion, a correct amplitude of density fluctuations, and an almost 
Harrison-Zel'dovich spectrum ($n_{S} \approx 0.96$); other tests as evading excess of non-Gaussianities and large tensor-to-scalar amplitude of fluctuations are 
also important, for more details see e.g. ref. \cite{Hinshaw:2012fq}.   

Based on the above mentioned remarks about the dark components, in the present work we present two different unification approaches that were partially worked out by 
us in recent years and here we test them further and remark some of their properties. The first model unifies dark matter and dark energy at the most trivial manner, identifying both of them with a 
single, {\it dark fluid}.  This is presented in next section, "The dark fluid".   The second model accomplishes the dynamics of the three dark components with a single scalar field, whose 
standard quadratic potential is responsible for the inflationary behavior,  and later,  "dark matter" domination is achieved through the specific dynamics of  scalar field whose 
non-trivial kinetic term possesses a minimum.  Finally, dark energy is realized by adding a proper (but not the standard) cosmological 
constant to the model.   The later model is present in section "Non-standard scalar field unification". Last, we present our conclusions at the end of the manuscript.  

\section{The dark fluid} \label{dfaa}
As we mentioned, the dark components of the Universe are decomposed, within the standard model of cosmology, in dark matter and dark energy.   However, this is 
only a possibility that in fact has support from historical reasons, but there are more ways to understand the dark sector, and specifically, in a unified way.  
Perhaps the simplest unified description of dark matter and dark energy is given by the so-called {\it dark fluid}. It is defined  
in a first approximation as a barotropic perfect fluid with adiabatic speed of sound equals to zero \cite{Aviles:2011ak} (see also \cite{Luongo:2011yk}),

\begin{equation}
 c^2_s = 0.  \label{DFdef}
\end{equation}

Being the fluid barotropic, this last condition implies that its perturbations do not develop acoustic oscillations, and therefore they grow at all scales by gravitational 
instabilities, behaving as cold dark matter. Several extensions to this model can be found in the literature, see for example \cite{Balbi:2007mz,Xu:2011bp,Caplar:2012ed,Aviles:2012et}. 
Without lost of generality we can write the equation of state of the dark fluid as 

\begin{equation}
 P_d(\rho_d) = w_d(\rho_d) \rho_d,  \label{EoS}
\end{equation}
where we have factorized the equation of state parameter $w_d$, which is a function only of the energy density of the fluid. From $c_s^2 = (\partial P/\partial \rho)_s =
dP/d\rho$, equations (\ref{DFdef}) and (\ref{EoS}) imply that 
$w_d(\rho) + \rho_d dw_d (\rho_d)/d\rho_d = 0$, and then $w_d(\rho_d) = - C / \rho_d$ where $C$ is a constant and the negative sign has been chosen for future convenience. 
From now on we will denote with a subindex {\it d} the variables of this dark fluid. The pressure is then

\begin{equation}
 P_d = -C.
\end{equation}

Thus, although the dark fluid perturbations grow at all scales, it is allowed to have a non-zero pressure. 
Astrophysical observations constrain this value to be very small, $|P| \ll \rho_A$, where $\rho_A$ is the energy density of typical astrophysical scales where
dark matter has been detected. Usually, it is
assumed that dark matter is pressureless, but this is by not means necessary, for instance it 
could be the case that $|P| \sim \rho_{c0}$,  where  $\rho_{c0}$ is a typical cosmological  energy density scale at present, without getting in contradiction with observations. 
In fact, this is the entrance that leads us to consider the dark fluid to be dark energy as well as dark matter.

Now, let us consider a  homogeneous and isotropic Universe at very large scales whose geometry is described by the Friedmann-Robertson-Walker metric, 
and that is filled with standard model particles ($b$, $\gamma$, ...) and with the above-defined dark fluid. The evolution equations of such a
Universe are

\begin{equation}
 H^2 = \frac{8 \pi G}{3}(\rho_d + \rho_b + \rho_{\gamma}), \label{s1}
\end{equation}
\begin{equation}
 \rho_b' + 3H\rho_b =0, \label{s2}
\end{equation}
\begin{equation}
\rho_{\gamma}' + 4 H \rho_{\gamma} =0, \label{s3}
\end{equation}
and
\begin{equation}
\rho_d' + 3H(1+w_d)  \rho_d  = 0, \label{s4}
\end{equation}
where prime means derivative with respect to cosmic time and $H\equiv a'/a$ is the Hubble factor. Equations (\ref{s2}) and (\ref{s3}) give 
$\rho_b = \rho_{b0} a^{-3}$ and $\rho_{\gamma} = \rho_{\gamma 0} a^{-4}$, where a subindex $0$ stands for quantities evaluated at present time, and 
we have normalized the scale factor  to be equal to one today, $a_0 = 1$. Integration of equation (\ref{s4}) gives

\begin{equation}
\rho_d = \frac{\rho_{d0}}{1+\mathcal{K}}\left( 1+ \frac{\mathcal{K}}{a^3} \right), \label{rho_d}
\end{equation}
where we have defined the constant $\mathcal{K} = (\rho_{d0} - C)/C$. This expression is exactly what one expects for a unified fluid: 
it contains a constant piece that behaves as dark energy and a second term that decays with the third power of the scale factor, just as a dark matter 
component does.  

Now, the equation of state parameter of the dark fluid becomes
\begin{equation}
 w_d = - \frac{1}{1+ \mathcal{K} a^{-3}}.  \label{EOSdf2}
\end{equation}
and its pressure, expressed in terms of the constant $\mathcal{K}$ instead of $C$, is 
\begin{equation}
P_d =  -  \frac{\rho_{d 0}}{1 + \mathcal{K}}.  \label{rorcdm2}
\end{equation} 
In order to ensure the positivity of the energy density at all times, the constant $\mathcal{K}$ must be a positive number. This implies that the pressure is
negative, a quality that allows the dark fluid to accelerate the Universe, and as we have outlined above it could take values of the order of
the critical density $(\sim 3 H_0^2/  8 \pi G)$ without affecting the behavior of the dark fluid as dark matter in astrophysical scenarios. 

In the $\Lambda$CDM model the equation of state parameter of the total dark sector, 
$w_T$, defined by 
\begin{equation}
 w_T \equiv \frac{ \sum_a  w_{a} \rho_{a}}{\sum_a \rho_a}, \label{EOST}
\end{equation}
where the subindex $a$ runs over dark matter (DM) and cosmological constant ($\Lambda$), is given by 
\begin{equation}
 w_T = -\frac{1}{1+\frac{\Omega_{DM}}{\Omega_{\Lambda}} a^{-3}}. \label{EOST2}
\end{equation}
(Note that in our convention $\Omega_i$ refers to the $i$-component abundance evaluated at present time.) 
Comparing these results to equations (\ref{s1}), (\ref{rho_d}), and (\ref{EOSdf2})  we note that under the identifications 
\begin{equation}
 \mathcal{K} = \frac{\Omega_{DM}}{\Omega_{\Lambda}}, \label{deg1rel}
\end{equation}
and
\begin{equation}
 \Omega_{d} = \Omega_{DM} + \Omega_{\Lambda}, \label{deg2rel}
\end{equation} 
the resulting cosmological background evolution in both models is exactly the same.  
What we have shown is that the dark fluid model is indistinguishable from the $\Lambda$CDM model at the background level. In the next subsection  
we shall show that these ideas can be extended for a complete cosmological description. 

This property has been called {\it dark degeneracy} by Martin Kunz in \cite{Kunz:2007rk}. In fact, it is more general than for the single fluid case worked here:
any  collection of fluids whose total equation of state parameter is equal to  equation (\ref{EOST2}) and that do not interact with 
baryons and photons will behave exactly as the composed dark matter-cosmological constant fluid, leading to a degeneracy with the $\Lambda$CDM model.

\subsection{Cosmological perturbations}

Now, let us consider cosmological perturbation theory in the conformal Newtonian gauge, the metric is given by (for details and notation see \cite{Aviles:2011ak, Ma:1995ey})
\begin{equation}
 ds^2 = a^2(\tau)\big[ -(1+2 \Psi) d\tau^2 + (1- 2 \Phi) \delta_{ij} dx^i dx^j \,\big],
\end{equation}
where $\tau$ is the conformal time related to the cosmic time by $dt = a d \tau$. The hydrodynamical equations in Fourier space for the dark fluid, 
obtained from $\nabla_{\mu}T^{\mu\nu}=0$, are given by

\begin{eqnarray}
 \dot{\delta}_d &=& - (1+ w_d)(\theta_d - 3 \dot{\Phi}) + 3 \mathcal{H}  w_d \delta_d 
         - 3 \mathcal{H} \frac{\delta P_d}{\delta \rho_d} \delta_d,  \label{pertddf} \\
 \dot{\theta}_d &=& - \mathcal{H} \theta_d  + k^2 \Psi + \frac{\delta P_d / \delta \rho_d}{1+w_d} k^2 \delta_d -  k^2 \sigma_d,  \label{perttdf}
\end{eqnarray}
where $\delta$ is the density contrast, $\theta$ the divergence of the peculiar velocity, and $\sigma$ the scalar anisotropic stress. 
For baryons after recombination, when the coupling to photons can be safety neglected, the hydrodynamical equations are
\begin{eqnarray}
 \dot{\delta}_b &=& -\theta_b + 3 \dot{\Phi},  \label{pertdb} \\
 \dot{\theta}_b &=& - \mathcal{H} \theta_b  + k^2 \Psi.  \label{perttb}
\end{eqnarray}
The fluid equations are supplemented with the Einstein's equations
\begin{equation}
 k^2 \Phi = -4 \pi G a^2 \sum_i \rho_i \Delta_i, \label{Ppe}
\end{equation}
and 
\begin{equation}
 k^2( \Phi - \Psi) = 12 \pi G a^2 \sum_i (\rho_i + P_i) \sigma_i
\end{equation}
where the sum runs over all fluid contributions and 
\begin{equation}
 \Delta_i = \delta_i + 3 \mathcal{H} (1+w_i) \frac{\theta_i}{k^2} 
\end{equation}
is the rest fluid energy density \cite{Bardeen:1980kt}.

To solve these equations, we need to add information about the nature of the dark fluid. 
The barotropic condition implies that $\delta P = c^2_s \delta \rho$ and, after equation (\ref{DFdef}), thus $\delta P_d =0$, and because it is a perfect fluid, the anisotropic stress vanishes, 
$\sigma_d = 0$. Therefore, in the 
right-hand side (rhs) 
of equation (\ref{pertddf}) the last term vanishes, and in the rhs of equation (\ref{perttdf}) only the two first terms survive. Moreover, if we solve only for baryons and the 
dark fluid the two gravitational potentials are equal, $\Psi = \Phi$.

\begin{figure}
\label{fig1}
\begin{minipage}{6.5in}
\begin{center}
\includegraphics[width=3in]{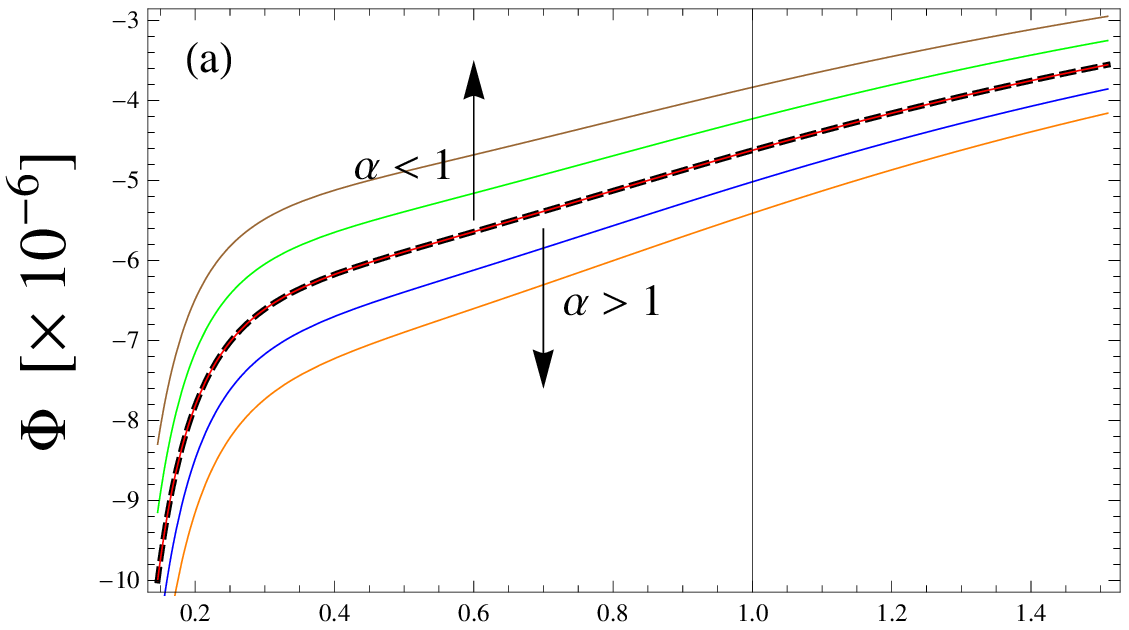} 
\includegraphics[width=3in]{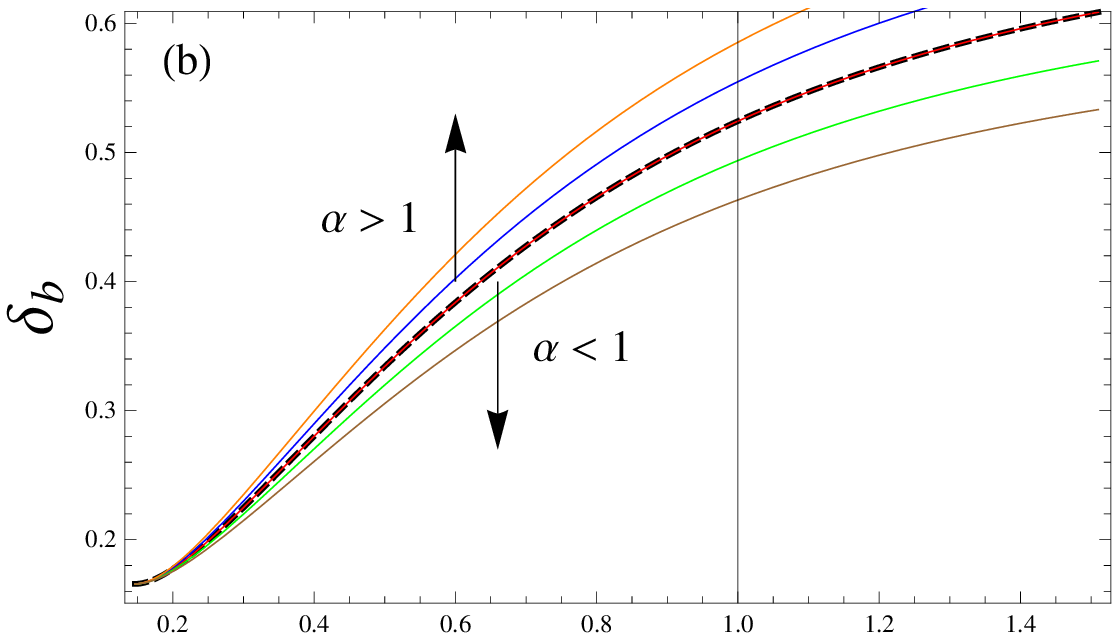} \\ [1\baselineskip]
\includegraphics[width=3in]{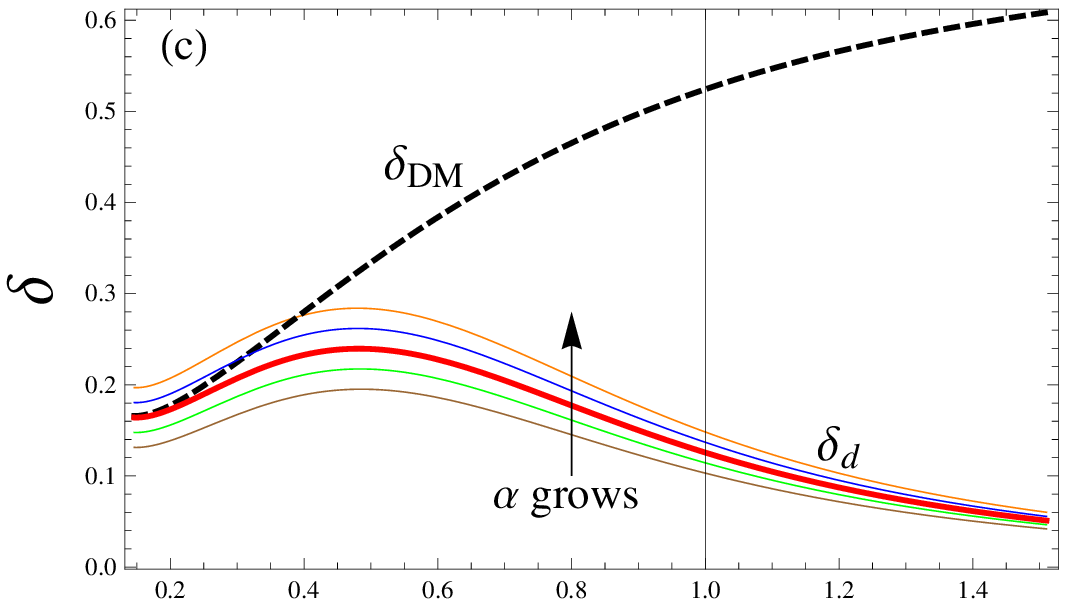} 
\includegraphics[width=3in]{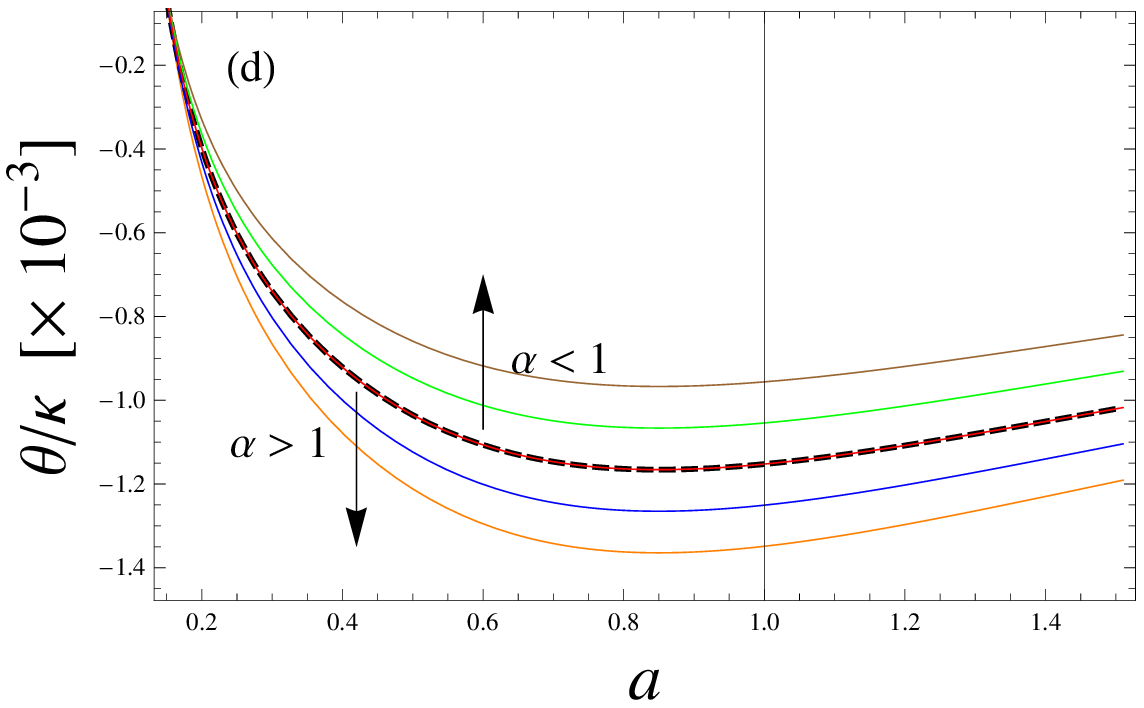}
\caption{Evolution of perturbation variables for a mode $k = 0.05\,\rm{Mpc}^{-1}$. 
         Solid curves are obtained from the dark fluid model for different values of the parameter $\alpha$ in the initial conditions 
          $\delta_d(\tau_i) = \alpha \rho_{\rm{DM}}(\tau_i) \delta_{\rm{DM}}(\tau_i) / \rho_d(\tau_i)$. $\alpha$ takes  values from 0.8 to 1.2.
          Dashed curves are for the $\Lambda$CDM model variables. The panels show: 
          (a) Gravitational potential $\Phi$. (b) Baryonic density contrast $\delta_b$. (c) Dark fluid (solid lines) 
          and dark matter (dashed line) density contrasts, $\delta_{d}$ and  $\delta_{\rm{DM}}$. (d) Dark fluid (solid lines)  
          and dark matter (dashed line) velocities, $\theta_d$ and $\theta_{\rm{DM}}$.  The solutions for the case $\alpha =1$ are depicted with the thick  (red) lines, 
          which for panels (a), (b), and (d) coincide with the dashed lines.}
\end{center}
\end{minipage}
\end{figure}

Figure \ref{fig1} shows the evolution of  a mode $k = 0.05\, \rm{Mpc}^{-1}$ of the perturbations variables.
The dashed lines shows results for the $\Lambda$CDM model, for which we have used the standard dark matter hydrodynamical perturbations equations
instead of equations (\ref{pertddf}) and (\ref{perttdf}). 

In evolving the perturbations for both models, $\Lambda$CDM and dark fluid, we have imposed on the initial conditions the relations  $\delta_d(\tau_i) = \alpha  \rho_{\rm{DM}}(\tau_i) 
\delta_{\rm{DM}}(\tau_i) / \rho_d(\tau_i)$ for the density contrasts, and $\theta_d(\tau_i) = \alpha  \rho_{\rm{DM}}(\tau_i) \theta_{\rm{DM}}(\tau_i) / (1+w_d)\rho_d(\tau_i)$ 
for the velocities, and we let $\alpha$ to take different values. These initial conditions are given at 
an initial time  well after recombination, so the relation $\delta_{\rm{DM}} \simeq \delta_b$ holds and we can neglect the coupling between baryons and photons. 

If we take $\alpha=1$ we note the evolution of the baryonic density contrast is undistinguished in both models. Then, although the cosmological observable is 
the baryonic matter power spectrum which includes a wide range of wavelengths, Figure 1 suggests that indeed the results are the same for both models for any wavelength. 
In fact this is true and it is shown in \cite{Aviles:2011ak}.  In the cosmological  context, imposing these two initial 
conditions is equivalent to demand  
that at first order in perturbation theory, the time-time and time-space components of the perturbed energy momentum tensor of the dark fluid and $\Lambda$CDM models
are equal  at the given initial time $\tau_i$. By the fact that we 
are using General Relativity which is a theory  with a well posed Cauchy problem, it is implied that the conditions will be preserved at all times; thus, equations 
\begin{equation}
 \rho_d \delta_d = \rho_{\rm{DM}} \delta_{\rm{DM}}, \label{ass1}
\end{equation}
and
\begin{equation}
 \rho_d(1+w_d) \theta_d = \rho_{\rm{DM}} \theta_{\rm{DM}}.  \label{ass2}
\end{equation}
hold at anytime. 

This analysis shows that the degeneracy between the dark fluid and the $\Lambda$CDM model is preserved at the linear cosmological order. 
To go beyond the linear order, let us make  perturbation expansions to the dark fluid  and $\Lambda$CDM
energy  momentum tensors about the (zero order) background cosmological fluids as 
\begin{equation}
 T_{\mu\nu} =  T_{\mu\nu}^{(0)} + T_{\mu\nu}^{(1)} +T_{\mu\nu}^{(2)} + \cdots. \label{EMTexp}
\end{equation}
If the total energy momentum tensors of both models are equal $\,\,\big(T_{\mu\nu}^{\,d} = T_{\mu\nu}^{\Lambda\rm{CDM}}\big)$, clearly each of the terms 
in the expansion will be equal as well $\big(T_{\mu\nu}^{\,d\,(i)} = T_{\mu\nu}^{\Lambda\rm{CDM}\,(i)} \big)$. This  argument is correct and it is  outlined in \cite{Kunz:2007rk}
to argue that the degeneracy is preserved at all orders in perturbation theory.  

Nonetheless, we want to stress a different approach: We are affected gravitationally by the total energy momentum tensor, but usually
when comparing observations to models we expand it as in equation (\ref{EMTexp}), and after this we assign values to each of the pieces. 
The fact that both energy momentum tensors are equal, say at zero order, does not imply that they will be equal at first order. 
In this situation, equations such as (\ref{ass1}) and (\ref{ass2}) are conditions of the theory and not consequences of it, and if not imposed, 
the degeneracy is broken, as seen in Figure \ref{fig1} for $\alpha \neq 1$. 

\subsection{Interactions to baryons}

From the last subsection it is clear that the dark fluid, although not necessarily, could be the sum of a dark matter and a cosmological constant components. 
Nonetheless, if the interaction to the particles of the standard model is only gravitational (which is the ultimate definition of {\it dark}), 
the nature of the dark fluid is fundamentally impossible to elucidate, because of the universality of this force. 
In this subsection we explore the possibility that the interactions between the dark fluid and baryons make the two models distinguishable.
The conservation of the energy momentum tensor is
\begin{equation}
 \nabla_{\mu} T^{\mu\nu}_{a} = Q^{\nu}_{a}, \label{Biid}
\end{equation}
where the energy momentum transfer vectors, $Q^{\nu}_{a}$, obey the constraint $ \sum_a Q^{\nu}_{a} = 0$,
and the sum runs over baryons and the dark fluid components.\footnote{In this work we will not consider interactions to electromagnetism. This is not only for simplicity,
many theoretical models present conformal couplings, such as the chameleon theories \cite{Khoury:2003aq,Khoury:2003rn} 
(and in general, scalar tensor gravity theories \cite{Brans:1961sx}), or even direct couplings
to the trace of the energy momentum tensor \cite{Sami:2002se,Aviles:2010ui,Hui:2010dn}. 
Also, this is expected in scenarios like the strong interacting dark matter  \cite{Spergel:1999mh,Wandelt:2000ad}, where the couplings are given through the strong force.} If we
consider the background continuity equations to be $\dot{\rho}_a + 3  \mathcal{H} (1+w_a) \rho_a = q_a$, it follows that up to first order cosmological perturbation theory 
in conformal Newtonian gauge
\begin{eqnarray} 
 Q^{0} &=&  \frac{1}{a^2} \big( q(1-\Psi) + \delta q \big)  \label{Q0} \\
  Q^{i} &=& \frac{1}{a^2} q v^i + \frac{1}{a^2} f^{,i} + \frac{1}{a^2} \epsilon^i. \label{Qi}  
\end{eqnarray}
Note that we have defined $\epsilon^i$ as a transverse vector and then it does not enter into the scalar perturbation equations.

We consider models in which the background cosmology is the same as in the $\Lambda$CDM model, 
accordingly we do not allow energy transfer ($q=0$) between the cosmic components. Nevertheless, we allow a momentum transfer different from zero. Thus, the interactions
affect the fluids only at first order in perturbation theory. The hydrodynamical equations for the perturbations are \cite{Aviles:2011ak}, for the dark fluid,
\begin{eqnarray}
 \dot{\delta}_d &=& - (1+ w_d)(\theta_d - 3 \dot{\Phi}) + 3 \mathcal{H}  w_d \delta_d + \frac{\delta q_d}{\rho_d}
                              \label{pertddfc} \\
 \dot{\theta}_d &=& - \mathcal{H} \theta_d  + k^2 \Psi - \frac{k^2 f_d}{\rho_d (1+w_d)},  \label{perttdfc}
\end{eqnarray}
and for baryons
\begin{eqnarray}
 \dot{\delta}_b &=& -\theta_b + 3 \dot{\Phi} + \frac{\delta q_b}{\rho_b},  \label{pertdbc} \\
 \dot{\theta}_b &=& - \mathcal{H} \theta_b  + k^2 \Psi + c^2_{sb} k^2 \delta_b - \frac{k^2 f_b}{\rho_b (1+w_b)}.  \label{perttbc}
\end{eqnarray}
For brevity, we have omitted the interactions of baryons to electromagnetism in the last equation.

In the absence of a fundamental theory we parametrize the coupling with
\begin{equation}
\delta q_d= 0, 
\end{equation}
and
\begin{equation}
 f_{d} = \rho_d(1+w_{d})  \frac{(\Sigma_{I} + \Sigma_{II}a^2)\rho_{d0}}{m_p a^2}   (\theta_{b} - \theta_d) / k^2,  \label{interaction3}
\end{equation}
where the parameters $\Sigma_{I}$ and $\Sigma_{II}$ have  units of area times velocity, or thermalized cross section $\langle \sigma v \rangle $, which we identify with  some, unknown, 
fundamental interaction. $n_d$ is the number density of  dark {\it particles} that we set equal to 
\begin{equation}
 n_d= \frac{\rho_{d0}}{m_p a^{3}},
\end{equation}
where we use $m_p$, the mass of the proton, as an arbitrary mass scale and $\rho_{d0}$ is the energy density of the dark fluid evaluated today. Here, we have not an 
analogous to the ionization fraction, in empathy to universal interactions. 
The first interaction, $\Sigma_{I}$, in equation (\ref{interaction3}) is inspired by electromagnetism while the second, $\Sigma_{II}$, by chameleon theories.

To constrain the interactions, we perform a Monte Carlo Markov Chain (MCMC) analysis over the nine-parameter space (Model A) 
$\{ \Omega_b h^2, \Omega_{\rm{DM}} h^2, \theta, \tau, n_s, \log A_s, A_{sz}, \Sigma_I, \Sigma_{II} \}$ using the code CosmoMC \cite{Lewis:2002ah}. 
The primordial scalar perturbations amplitude $A_s$  is given at a pivot scale of  
$k_0 = 0.05\, \rm{Mpc}^{-1}$. 

We have imposed flat priors on the two interaction parameters: $0<\Sigma_I< 10^{-7 } \times \sigma_T$ and $-11 \times \sigma_T <\Sigma_{II}< 10 \times \sigma_T$. 
For the CMB anisotropies and polarization data
we used the Wilkinson Microwave Anisotropy Probe (WMAP) seven-year observations results \cite{Larson:2010gs}. For the joint analysis we use also Hubble  
Space Telescope measurements (HST) \cite{Riess:2009pu} to impose a Gaussian prior on the Hubble constant today of $H_0 = 74 \pm 3.6 \,\rm{km/s/Mpc}$, 
and the supernovae type Ia Union 2 data set compilation by the Supernovae Cosmology Project \cite{Amanullah:2010vv}, we have named these three different observations 
as Set I, because this is the one used in \cite{Aviles:2011ak}. Additionally, here we use the catalog of luminous red galaxies SDSS DR7 LRG given in \cite{Reid:2009xm}. 
It is worth noting that this method represents only a rough estimate of the parameters because of the galactic bias problem.

We also study two other models: Model B, only considering the interaction $\Sigma_I$, it has an eight-parameter space  
$\{ \Omega_b h^2, \Omega_{\rm{DM}} h^2, \theta, \tau, n_s, \log A_s, A_{sz}, \Sigma_I \}$; 
and Model C, which does not consider any interaction, a seven-parameter space  
$\{ \Omega_b h^2, \Omega_{\rm{DM}} h^2, \theta, \tau, n_s, \log A_s, A_{sz} \}$, corresponding to the standard $\Lambda$CDM model.

The summary of constraints is outlined in Table \ref{table:nonlin}. In Figure 2 we show the contour 
confidence intervals for the  marginalized $\Sigma_I\!-\!\Sigma_{II}$ space at 0.68 and 0.95 confidence levels (c.l.). There,  the
high degeneracy between both parameters is shown: while $\Sigma_{II}$ takes values closer to zero, $\Sigma_I$ also does.
It is interesting that nonzero values of the interactions (when introduced) are consistent and preferred by the considered data at 0.95 c.l. when using
Set I of observations only. When including the SDSS DR7 LRG data, $\Sigma_I$ and $\Sigma_{II}$ include the zero at 
1$\sigma$ and 2$\sigma$ c.l., respectively.

\begin{table}
\caption{Summary of constraints. The upper panel contains the parameter spaces explored with MCMC for each one of the three models. 
The bottom panel contains derived parameters. The data used are Set I of observations: WMAP seven-year data, Union 2 compilation and HST, which are the 
used in \cite{Aviles:2011ak}; and the catalog SDSS DR7 LRG.   
}          
\begin{tabular}{l|c|c|c}
\hline\hline
{\small Parameter}   & {\small Model A$\,{}^{a}$ }& {\small Model B$\,{}^{a}$} & {\small Model C$\,{}^{a}$} \\ [1.5ex]
                 
\hline 

 {\small $10^2 \Omega_b h^2$}         & {\small $2.247$}{\tiny${}_{-0.047}^{+0.047}$}         & {\small $2.264$}{\tiny ${}_{-0.054}^{+0.057}$}        
                                      & {\small $2.270$}{\tiny ${}_{-0.054}^{+0.054}$} \\[0.8ex] 
{\small $\Omega_c h^2$}               & {\small $0.1201$}{\tiny${}_{-0.0040}^{+0.0041}$}      & {\small $0.1147$}{\tiny ${}_{-0.0030}^{+0.0030}$}      
                                      & {\small $0.1133$}{\tiny ${}_{-0.0031}^{+0.0031}$} \\[0.8ex]
{\small $\theta$}                     & {\small $1.043$}{\tiny${}_{-0.003}^{+0.003}$}         & {\small $1.040$}{\tiny ${}_{-0.002}^{+0.002}$}        
                                      & {\small $1.040 $}{\tiny ${}_{-0.003}^{+0.003}$} \\[0.8ex]
{\small $\tau$}                       & $\quad${\small$0.08733$}{\tiny${}_{-0.00680}^{+0.00580}$}$\quad$    & $\qquad${\small $0.08879$}{\tiny ${}_{-0.0068}^{+0.0055}$}$\quad$  
                                      & {\small$0.08797$}{\tiny${}_{-0.00627}^{+0.00618}$}  \\[0.8ex]
{\small $10^{8} \Sigma_I\,{}^b$}              & {\small $0.344$}{\tiny${}_{-0.244}^{+0.205}$}         & {\small $0.0894$}{\tiny ${}_{-0.0400}^{+0.0493}$}  
                                      & $--$  \\ [0.8ex] 
{\small $\Sigma_{II}\,{}^b$}                & {\small $-3.133$}{\tiny ${}_{-1.402}^{+1.448}$}       & $--$                             
                                      & $--$ \\[0.8ex]
{\small $n_s$}                        & {\small $0.9813$}{\tiny ${}_{-0.0159}^{+0.0156}$}     & {\small $0.9658$}{\tiny ${}_{-0.0134}^{+0.0130}$} 
                                      & {\small $0.9686$}{\tiny ${}_{-0.0124}^{+0.0125}$} \\[0.8ex]
{\small $\log[10^{10} A_s]$}$\quad$   & {\small $3.134$}{\tiny ${}_{-0.041}^{+0.040}$}        & {\small $3.095$}{\tiny ${}_{-0.032}^{+0.030}$} 
                                      & {\small $3.089$}{\tiny ${}_{-0.033}^{+0.032}$} \\[0.8ex]
{\small $A_{SZ}\,{}^c$}               & {\small $1.091 \pm 0.563$}                            & {\small $0.872 \pm 0.562$} 
                                      & {\small $0.927 \pm 0.564$} \\ [1ex]
\hline 
{\small $\Omega_d$}          & {\small $0.951$}{\tiny ${}_{-0.023}^{+0.023}$}              & {\small $0.952$}{\tiny ${}_{-0.021}^{+0.021}$} 
                             & {\small $0.953$}{\tiny ${}_{-0.020}^{+0.019}$}\\ [0.8ex] 
{\small $\mathcal{K}$}       & {\small $0.373$}{\tiny ${}_{-0.037}^{+0.038}$}              & {\small $0.336$}{\tiny ${}_{-0.029}^{+0.030}$}   
                             & {\small $0.322$}{\tiny ${}_{-0.028}^{+0.029}$} \\ [0.8ex]
{\small $t_0$}               & {\small $13.74$}{\tiny ${}_{-0.12}^{+0.11}\,$} {\small Gyr} & {\small $13.79$}{\tiny ${}_{-0.11}^{+0.11}\,$} {\small Gyr} 
                             & {\small $13.77 $}{\tiny ${}_{-0.11}^{+0.12}\,$}{\small Gyr} \\ [0.8ex]
{\small $\Omega_{\Lambda}$}  & {\small $0.693$}{\tiny ${}_{-0.019}^{+0.019}$}              & {\small $0.713$}{\tiny ${}_{-0.016}^{+0.015}$} 
                             & {\small $0.721$}{\tiny ${}_{-0.016}^{+0.016}$}\\[0.8ex]
{\small $H_0\,{}^d$}         & {\small $68.15$}{\tiny${}_{-1.34}^{+1.35}$}                 & {\small $69.22$}{\tiny ${}_{-1.40}^{+1.39}$} 
                             & {\small $69.89$}{\tiny ${}_{-1.46}^{+1.32}$} \\ [0.8ex]
\hline
\end{tabular}


\label{table:nonlin} 
\end{table}

\begin{figure}
\begin{minipage}{6.5in}
\begin{center}
\includegraphics[width=2.4in]{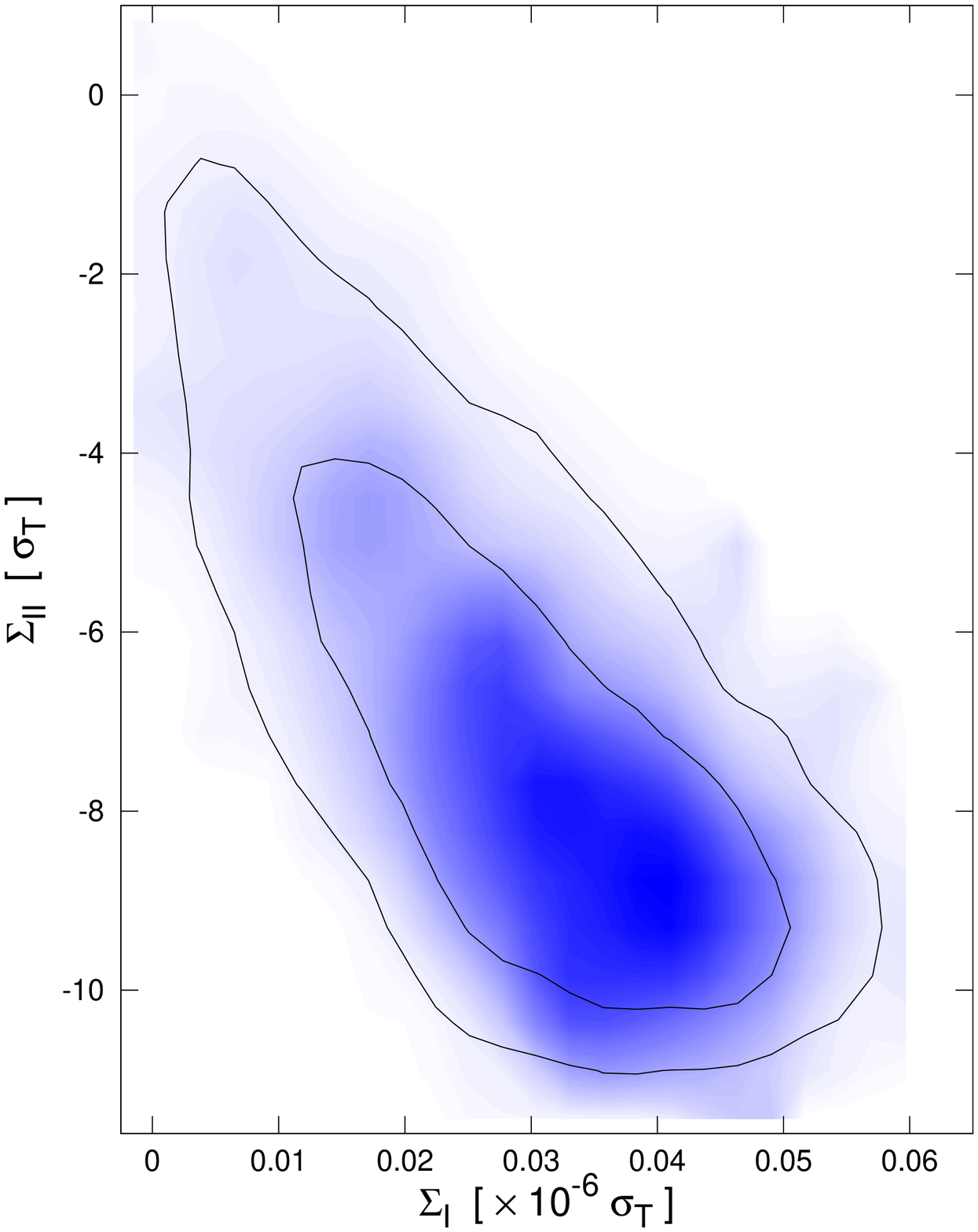} 
\includegraphics[width=2.3in]{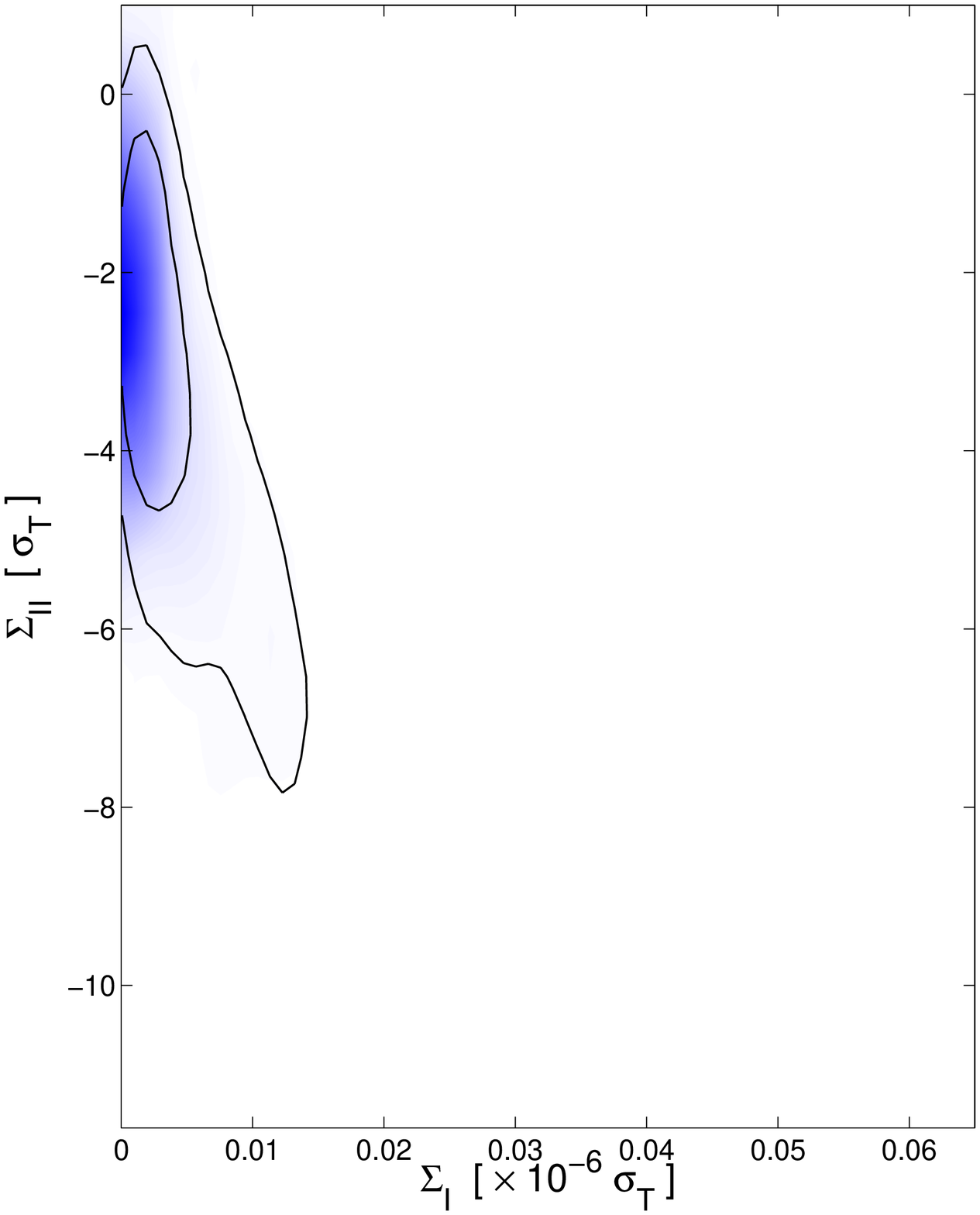}
\caption{Contour confidence intervals for $\Sigma_I$ vs $\Sigma_{II}$ at $68 \%$ and $95 \%$ c.l.  The shading shows the mean likelihood of the samples. 
Left panel: Considering Set I of observations only. Right panel: Considering Set I and SDSS DR7 LRG observations.}
\end{center}
\end{minipage}
\label{fig4}
\end{figure} 

Instead of using the proton mass as the scale in the interactions, we can use an  arbitrary associated mass to the dark fluid  ``particles'', $m_d$. 
We obtain the following constraints at 0.68 c.l. on the ratio $\Sigma/m_d$ (we use $c = 3 \times 10^{10} \rm{cm/s}$):

For the case in which we consider both interactions (Model A)

\begin{equation}
0.21 \times 10^{-22} <  \frac{\Sigma_I}{m_d} < 1.17 \times 10^{-22} \,\,\frac{\rm{cm}^3/\rm{s}}{\rm{GeV}/c^2}
\end{equation}
and
\begin{equation}
-0.96  \times 10^{-13}  <  \frac{\Sigma_{II}}{m_d} <  -0.35 \times 10^{-13} \,\, \frac{\rm{cm}^3/\rm{s}}{\rm{GeV}/c^2}.
\end{equation}
While for Model B,
\begin{equation}
0.10 \times 10^{-22} <  \frac{\Sigma_I}{m_d}<  0.29 \times 10^{-22} \,\, \frac{\rm{cm}^3/\rm{s}}{\rm{GeV}/c^2}. 
\end{equation}

Finally, in Figure 3 we show the plots of the angular and matter power spectrums using the best fits values obtained by the MCMC fitting procedure and shown in Table 1. 

\begin{figure}
\begin{minipage}{6.5in}
\begin{center}
\includegraphics[width=2.8in]{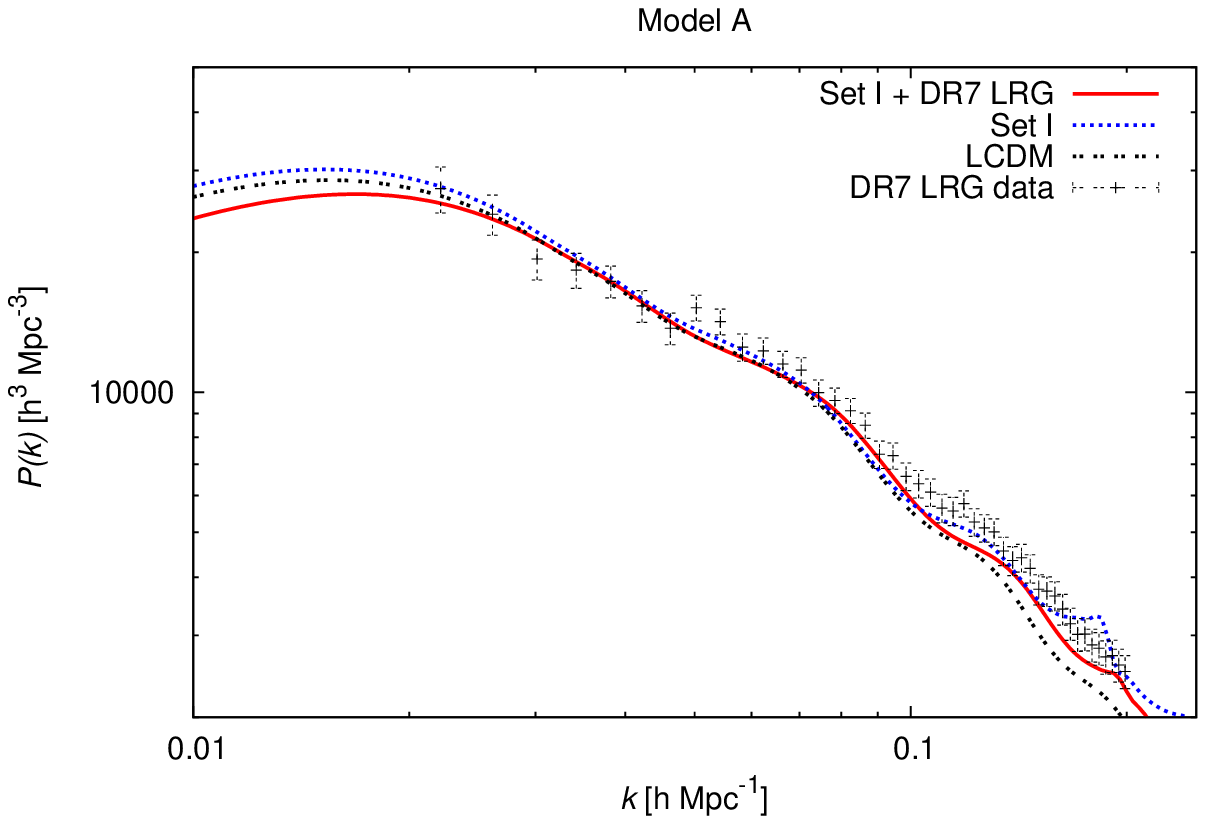} 
\includegraphics[width=2.8in]{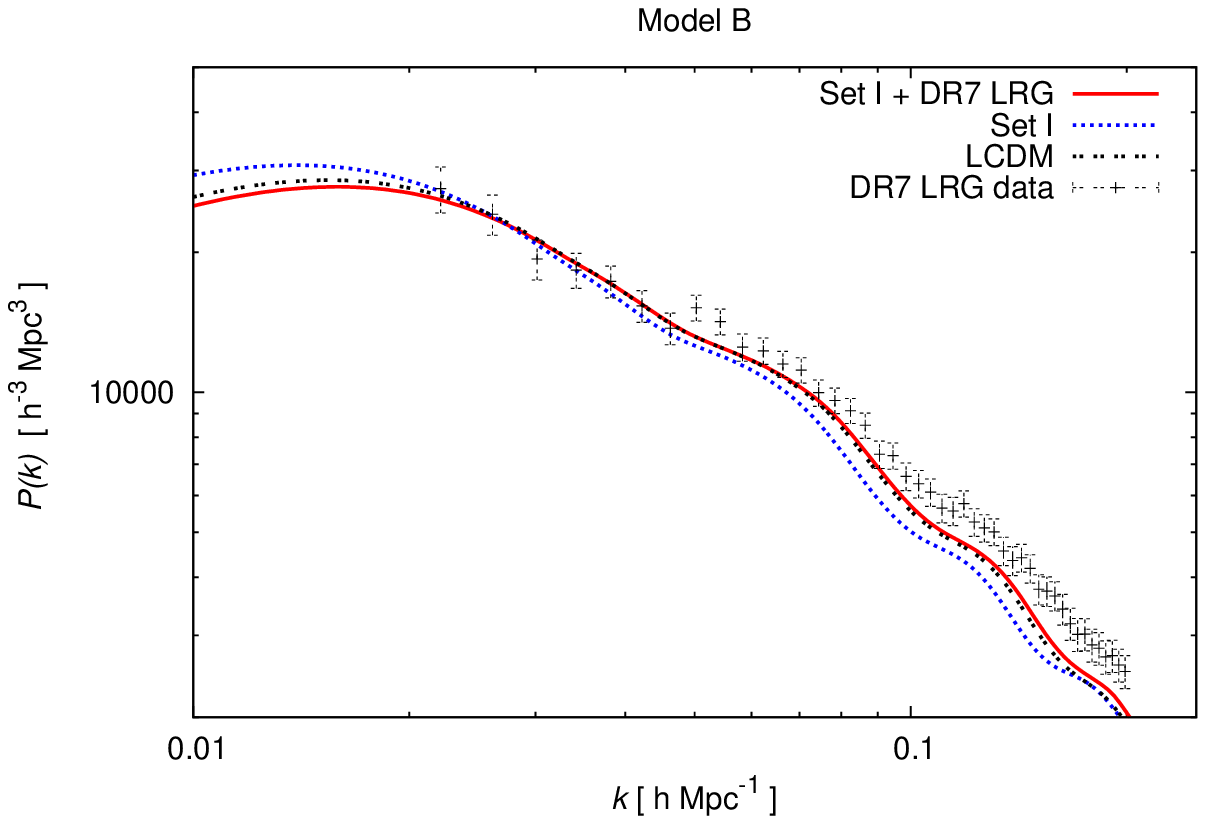} \\ [1\baselineskip]
\includegraphics[width=3in]{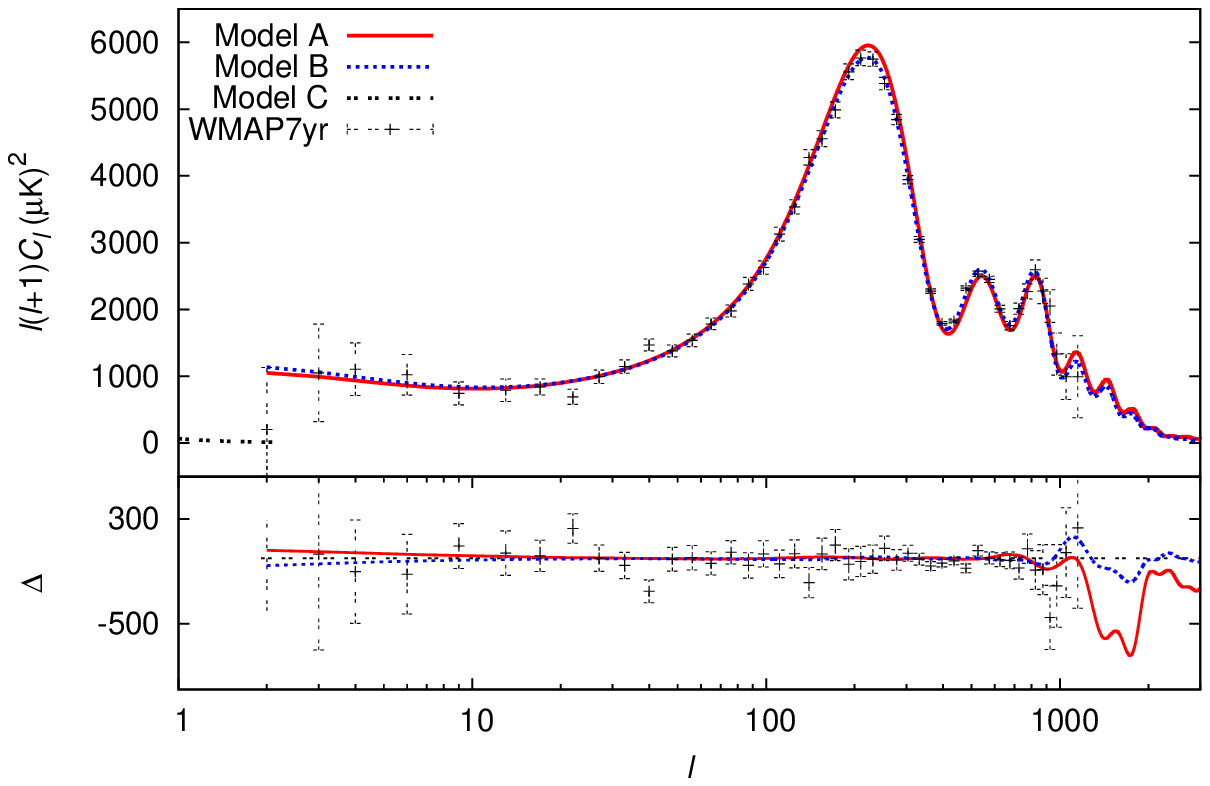} 
\caption{Matter power spectrums for Model A and Model B at redshift $z=0$ and CMB power spectrum for the three models. The parameter values are taken from Table 1.}
\end{center}
\end{minipage}
\label{fig5}
\end{figure} 

Note that we have not considered the effect that interactions $I$ and $II$ could have on big bang nucleosynthesis. This is because our 
phenomenological model only includes the thermalized cross sections $\Sigma_I$ and $\Sigma_{II}$ of elastic collisions, whose strengths 
are at least 9 orders of magnitude weaker than the Thomson scattering at this epoch,  and more important, the interactions do not annihilate 
baryons and, therefore, maintain the baryon-to-photon ratio unaltered.  Accordingly,  we expect the effect over this process to be quite weak. 


\section{Non-standard scalar field unification} \label{scalar-field-model}

We now turn to another theoretical scheme that pursues to unify the dark fluids, and it is through a scalar field. There has been works trying 
to unify dark matter, dark energy, and inflation using scalar fields. In the work in ref.  \cite{Liddle:2006qz}, where a
single scalar field with quadratic potential is used to produce the three phenomena, the inflation phase is driven
by the potential as in the usual chaotic inflation scenario. An incomplete
reheating phase leaves enough energy in the field for it to oscillate around the
minimum of the potential and behave as dark matter \cite{Magana:2012ph},
while a constant term in the potential allows it to reproduce dark energy.  However, as mentioned in the Introduction, a fine tuning of the parameters is 
necessary to accommodate the proper dark matter content. 

Other works have used a generalized version of the scalar field Lagrangian in order
to accomplish the unification. There the Lagrangian has the form
\begin{equation}\label{general_lagrangian}
 \lag = \lag (X,\phi) ,
\end{equation}
where the $X=-\frac{1}{2}\phi_{,\mu} \phi^{,\mu}$ is the usual kinetic term, and the
Lagrangian is a general function of it and the scalar field $\phi$. This type of scalar
fields have been used to model inflation \cite{ArmendarizPicon:1999rj},
dark energy \cite{Chiba:1999ka}, and also to unify dark energy
and dark matter \cite{Chimento:2003ta, Scherrer:2004au,Bertacca:2010ct}. Combining 
one Lagrangian proposed in \cite{Chimento:2003ta} with an appropriate potential term
it is possible to obtain a unification of inflation, with dark energy and dark matter
as it was made in \cite{Bose:2008ew,Bose:2009kc,DeSantiago:2011qb}.

\subsection{Conditions for Dark Energy and Dark Matter unification}

A general Lagrangian of the form of equation (\ref{general_lagrangian}) has an associated energy density
\begin{equation}
 \rho = 2X\lag_{,X} - \lag,
\end{equation}
and pressure
\begin{equation}
 P=\lag.
\end{equation}
From that one can obtain the equation of state
\begin{equation}
 \omega = \frac{\lag}{2X\lag_{,X} - \lag},
\end{equation}
and sound speed $c_s^2=P_{,X}/\rho_{,X}$ given by \cite{Garriga:1999vw}
\begin{equation}
 c_s^2 = \frac{\lag_{,X}}{2X\lag_{,XX}+\lag_{,X}}\,.
\end{equation}

If the Lagrangian is the sum of a constant that accounts for the dark energy
and a variable part that accounts for the dark matter $\lag = \lag_0 + \lag_{(dm)}(X,\phi)$, the conditions on the
latter are that $\omega_{dm} \ll c^2$ and ${c_{(dm)}}_s^2 \ll c^2$, as mentioned in the Introduction. The first condition is in order to have a
background evolution similar to that of $\Lambda$CDM and the second one in order
to allow for structure formation; however this condition can be violated in some models
and still have structure formation \cite{Magana:2012ph}. The conditions on the 
Lagrangian become
\begin{equation}\label{dm1}
 \frac{\lag_{(dm)}}{X\lag_{(dm),X}} \ll 1 \,, 
\end{equation}
and
\begin{equation}\label{dm2}
  \frac{\lag_{(dm),X}}{X\lag_{(dm),XX}}\ll 1 \,.
\end{equation}
These conditions are satisfied by a Lagrangian with a minimum at an $X\ne 0$ and the field close to that minimum, so the standard kinetic term $F=X$ does not fulfill them. From the
first condition the value of the dark matter part of the Lagrangian in the minimum should be zero.

In the work by Scherrer \cite{Scherrer:2004au} the Lagrangian
\begin{equation}
 \lag =	F_0 + F_m(X-X_0)^2
\end{equation}
is used, where the first term is a constant that accounts for the dark energy and the second term
behaves as dark matter as it satisfies equations (\ref{dm1}, \ref{dm2}). It is argued in ref. \cite{Giannakis:2005kr} that this model changes the 
transfer function, and it is concluded that in order to account for the $\Lambda$CDM power spectrum the deviation from the minimum 
$\epsilon \equiv (X-X_0)/X_0$ should be smaller than $10^{-16}$ in the present epoch. Fulfilling this condition guarantees this model to be indistinguishable 
from cold dark matter perturbation growth.  

Let us now consider another model. In a previous work  \cite{DeSantiago:2011qb}, we employed  a Lagrangian 
proposed in ref. \cite{Chimento:2003ta} with an extra constant term form \cite{Bose:2008ew}:
\begin{equation}\label{lagrangian2}
	\lag = \frac{1}{(2\alpha-1)} \left[ (A X)^{\alpha}  
	    - 2\alpha  \alpha_0  \sqrt{AX} \right] + M.
\end{equation}
Here the effective constant term that accounts for dark energy is
\begin{equation}
 \lag_0 = M-\alpha_0^{2\alpha/(2\alpha-1)} \,,
\end{equation}
and the (dark matter) part that satisfies equations (\ref{dm1}, \ref{dm2})
\begin{equation} \label{lag_dm}
	\lag_{dm} = \frac{1}{(2\alpha-1)} \left[ (A X)^{\alpha}  
	    - 2\alpha  \alpha_0  \sqrt{AX} \right] + \alpha_0^{2\alpha/(2\alpha-1)}.
\end{equation}
The conditions for this Lagrangian to satisfy the cosmological constraints were studied first in ref. 
 \cite{Bose:2008ew} for the $\alpha=1$ case and later in ref. \cite{DeSantiago:2011qb} for the general case with
$n=2\alpha/(2\alpha-1)$ a positive integer. The mathematical advantage of this Lagrangian is that the 
cosmological evolution of the energy density in a flat Friedmann-Robertson-Walker Universe is reduced
to the simple equation
\begin{equation}
 \rho = \left[ \alpha_0 + \frac{c_0}{a^3} \right]^n - M \, .
\end{equation}
This expression can be split into a dark energy term, a dark matter term and extra terms that are functions of larger
powers of the scale factor, as follows
\begin{eqnarray}
   \rho_{de} &=& \alpha_0^n - M  ,\label{c} \\
   \rho_{dm} &=& \frac{nc_0\alpha_0^{n-1}}{a^3}   ,
\label{d}\\ 
   \rho_{extra} &=& \sum_{k=2}^n \binom{n}{k}\alpha_0^{n-k}\left(\frac{c_0}{a^3}\right)^k  .
\end{eqnarray}

The conditions (\ref{dm1}) and (\ref{dm2}) are fulfilled in this model too, since around the minimum the "dark matter" Lagrangian, equation (\ref{lag_dm}), 
behaves as Scherrer's model, with $F_{m} = \frac{1}{4} A^2 \alpha \, \alpha_0^{(2\alpha - 4)/(2\alpha - 1)}$ and $F_0 = - \rho_{de} $, see equation (\ref{c}), where we already 
added the constant $M$.

In order to have $\rho_{extra}$ negligible during the known evolution of the Universe to avoid spoiling the standard cosmic dynamics at least from nucleosynthesis to today, 
one is forced to demand the following  condition
\begin{equation}\label{nn}
	M \gg  \rho_{de0}\left[ \frac{z_{\rm nuc}^{3n-4}}{(3 n)^n}
	\frac{\rho_{de0}}{\rho_{r0}} \right]^{1/(n-1)},
\end{equation}
in other words, $M$ has to be more than $10^{10}$ times bigger than the magnitude of the dark energy today but at
the same time, from equation (\ref{c}), it has to cancel almost exactly with $\alpha_0^n$ to yield the correct value of dark energy. Of course, 
this is a fine tuning of the model. 

\begin{figure}
 \includegraphics[width=110mm]{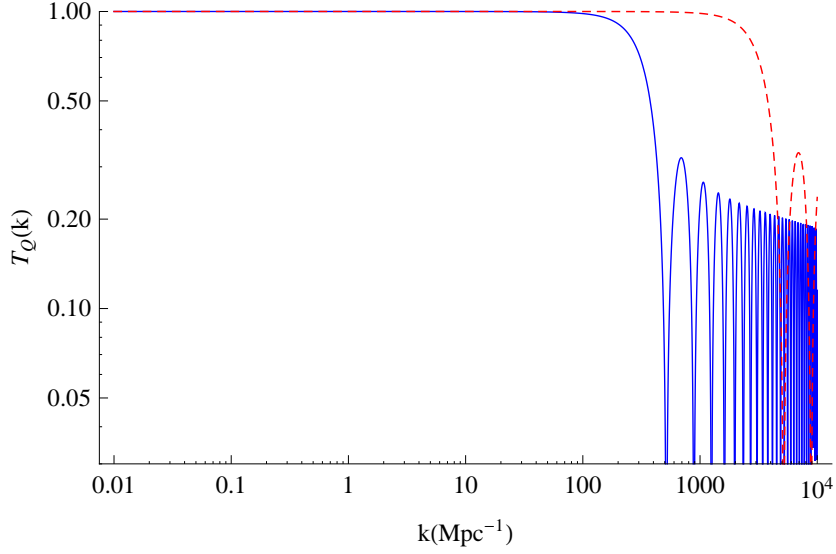}
  \caption{Deviation from the CDM transfer function for the cases $n=2$ (blue, continuos line) and $n=3$ (red, dashed line).
  The modification becomes significant only for small scales where linear theory
  is no longer valid.}
  \label{fig:trans}
\end{figure}

In the same way as in the Scherrer's model the field has to be close to the minimum, in order to have a correct transfer
function.  The condition given by equation (\ref{nn}) can be used to obtain a bound for the deviation $\epsilon=(X-X_0)/X_0$ as
\begin{equation}\label{deviation}
   \epsilon \ll (z+1)^3 2(n-1)(3n)^{1/(n-1)}10^{(-30n+36)/(n-1)}.
\end{equation}
which implies, for example, for $n=2$ that the value for $\epsilon$ today is smaller than $10^{-23}$ satisfying the condition
obtained from the transfer function in which $\epsilon < 10^{-16}$, see figure \ref{fig:trans}.

\subsection{Inflation}

So far the scalar field with Lagrangian given by equation (\ref{lagrangian2}) is able to reproduce the phenomena of dark matter and dark energy 
in the late Universe, and it only depends on the kinetic term $X$. If we add a potential term to the Lagrangian it can
account for the energy density during inflation. For this we chose in \cite{DeSantiago:2011qb} a quadratic potential
\begin{equation}\label{q-pot}
 V(\phi)=\frac{1}{2}m^2\phi^2 \,.
\end{equation}
Due to the large energies during inflation the kinetic term gets reduced to only the first term in the expression
(\ref{lagrangian2}) with $\alpha>0$, and with this simplification we obtain the effective Lagrangian at high energies
\begin{equation}
 \lag = \frac{1}{(2\alpha-1)} (A X)^{\alpha} - \frac{1}{2}m^2\phi^2 \,,
\end{equation}
which has been well studied as a source for inflation \cite{Mukhanov:2005bu,Panotopoulos:2007ky}. The slow roll (sr)
parameters for this non-canonical case are given by
\begin{eqnarray}
	\epsilon_{\rm sr} &=& \frac{M_{\rm pl}^2}{2F_X} \left( \frac{V'}{V} \right)^2,
	\label{epsilon}\\
	\eta_{\rm  sr} &=& \frac{M_{\rm pl}^2}{F_X^2} \frac{V''}{V},
	\label{eta}
\end{eqnarray}
where $F$ is the effective kinetic term in the Lagrangian and $F_X$ its derivative with respect to $X$.
Supposing that a slow roll regime holds during the inflationary epoch, the end of inflation occurs when it is
violated, and $\epsilon_{ \rm  sr} \sim 1$, which can be simplified to
\begin{equation}
	\phi_f^2 = \frac{2M_{\rm pl}^2}{F_X}.
	\label{vend}
\end{equation}
The beginning of inflation, can be calculated assuming 60 e-folds of inflation. Using the slow roll approximation
we find that $\phi_i = 15.5 M_{pl}$. The mass parameter $m$ can be computed in terms of the amplitude of perturbations
in the CMB as $m=7 \times 10^{-6} M_{pl} (n-1)^{1/4}$. With this data and using the fact that 
$F_X \sim 1$, we can compute the values of the slow roll
parameters at the beginning of inflation as $\epsilon_{ \rm  sr(i)} = 8.3\times 10^{-3}$ and $\eta_{ \rm  sr(i)} = 8.3\times 10^{-3}$.
And the spectral index and tensor to scalar ratio get the values
\begin{eqnarray}
 n_s &=& 1-0.3\sqrt{n-1} \,, \\
 r &=& 0.15 \sqrt{n-1} \,.
\end{eqnarray}
It yields a much redder power spectrum than Harrison-Zel'dovich's. The current measurement of the spectral index, from WMAP 9 years plus combined 
data from e-CMB, BAO, and $H_0$,  is \cite{Hinshaw:2012fq} $n_s=0.961 \pm 0.008$. To avoid this inconsistency, as the parameter $m$ and $\phi_i$ are constrained, the only possibility
is to adjust $F_X$ at the beginning of inflation, but we have not made this analysis yet.

\bigskip 

When considering all constraints for a successful  cosmological model, with exception to the reddish spectral index, 
the parameters have to comply with, for $\alpha=1$, 
\begin{eqnarray}
10^{-48} \mpl^2< &\alpha_0& <10^{-40}\mpl^2 \,, \nonumber \\ \nonumber
m &\sim& 10^{-6} \mpl \,, \\ \nonumber
\alpha_0^2 -M &\sim& 10^{-120} \mpl^4 \,, \\
A &\sim& 1 \,.
  \label{constants}
\end{eqnarray}
This would guarantee a cosmological dynamics that emulates that of the $\Lambda$CDM model over the 
whole expansion's history  and perturbed kinematics.  We notice that the energy scale of $\alpha_{0}^2$ and $M$ can be 
between $(10 \, {\rm keV})^4$ to $(100 \, {\rm MeV})^4$, but their difference must be very small to achieve the present cosmological constant value. 

\subsection{Phase space}

One may wonder how a transition from inflation to "matter" dominated happened and then to a dark energy dominated Universe, and how robust to the different initial conditions the system is. 
To answer this issue we have performed a phase space analysis of the solutions \cite{DeSantiago:2012mm} and we present an excerpt pointing out some features of the model. In ref. \cite{DeSantiago:2012nk} a study on the general features of the phase space for models
with Lagrangian $\lag =F(X)-V(\phi)$ is presented. Here
however, we will carry out a similar analysis adapted to the particular choice given  by  (\ref{lagrangian2},\ref{q-pot}). 

For concreteness, let us consider $\alpha=1$. It is straightforward to show that the system of first order 
autonomous equations becomes:  
\begin{eqnarray}
  \dot{z} &=& -\frac{m^2\phi}{\sqrt{A}} +
  \frac{\sqrt{3}}{2\mpl}\left( -\sqrt{2} z + 2 \alpha_0 \si (z) \right) \label{zdot}
  \sqrt{z^2+m^2\phi^2-2M} \,, \\
  \dot{\phi} &=& \frac{z}{\sqrt{A}} \,.
\end{eqnarray}
With the equation of state of the field $p_\phi/\rho_\phi$ written in terms of these variables as
\begin{equation}
  \omega_\phi = \frac{2M+z^2-\sqrt{8}\alpha_0|z|-m^2\phi^2}
  {-2M+z^2+m^2\phi^2} \,.
  \label{omegaphi}
\end{equation}

\begin{figure}[htbp]
  \includegraphics[width=100mm]{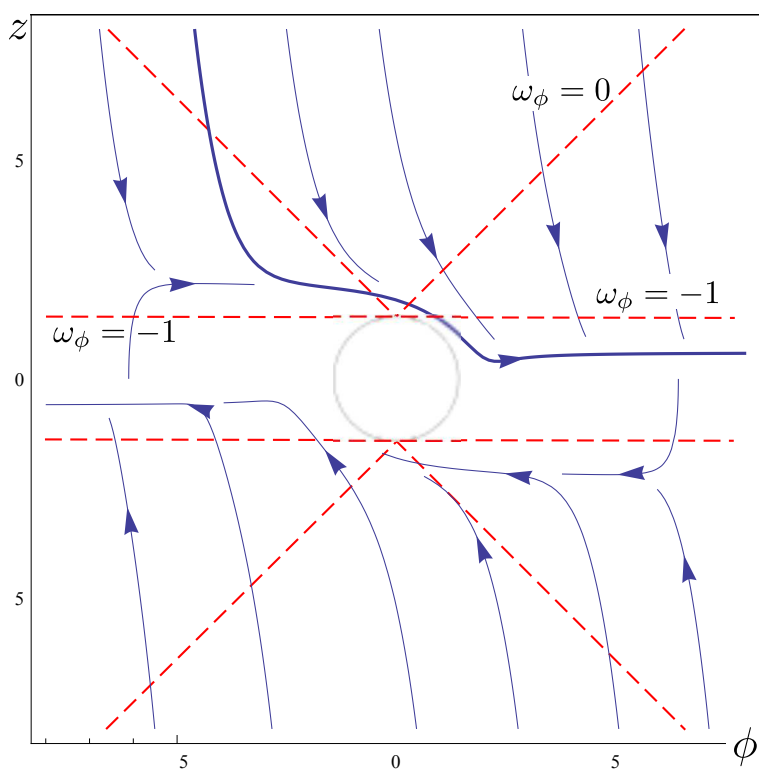}
  \caption{Phase space for the model, the continuous (blue) lines correspond to
  the evolution of the system. The dotted (red) lines correspond to lines of constant
  equation of state, both horizontal lines corresponding to $\omega_\phi=-1$ and
  the four $45^{\circ}$ segments correspond to $\omega_\phi=0$. A typical solution represented
  by the thick line approaches first to $\omega_\phi \sim -1$ (inflation),
  then passes through $\omega_\phi \sim 0$ emulating dark matter and finally comes back
  to $\omega_\phi \sim -1$ at late times as dark energy.}
  \label{fig:phase}
\end{figure}

The system doesn't have any critical points, but 
the system can be solved numerically to obtain its phase space, shown in Fig. 
\ref{fig:phase}. There we have plotted in dotted (red) lines those of constant equation of
state, the horizontal lines corresponding to $\omega_\phi=-1$ and diagonal lines to $\omega_\phi = 0$.  As can be
seen the sector of initial conditions with big negative $\phi$ values and positive
$z$ values evolves towards a solution with equation of state
near $-1$ which in the phase space corresponds to the left horizontal branch.
This in the unification models is interpreted as the initial period of inflation
in which the equation of state of the solution gets close to $-1$.

This solution later crosses the lines corresponding to a equation of state equal to 0  
(diagonal lines) that in the unification models correspond to the matter
domination epoch.
The time that the system stays in the regime of $\omega_\phi\sim 0$ has to be long
in order to mimic dark matter. This time will depend on the value of the parameters
(\ref{constants}) in the Lagrangian, and the parameters can be adjusted in accordance to equation (\ref{constants}) in order to obtain this
behaviour from a redshift of order $10^{10}$ up until a recent time when the transition to
$\omega_\phi <0$ has to occur. 
Finally, the solution evolves towards a second period of $\omega_\phi$
close to $-1$, that in the phase space corresponds to the right horizontal branch.
The whole behaviour occurs also for solutions beginning with big positive values of
$\phi$ and negative values of $z$, in which solutions go from positive to
negative values in $\phi$ and live in the $z<0$ part of the phase space, as can be
seen in Fig. \ref{fig:phase}.

We conclude that an important sector of the possible initial conditions can yield  
the expected behavior needed to unify the phenomena of dark matter, dark energy, and
inflation with a single, albeit complicated, scalar field. 

\section{Conclusions} \label{conclu}

The standard model of cosmology is built with standard particle physics and its interactions, and in addition, one has dark matter and dark energy. 
On the other hand, an inflationary dynamics is also necessary and it demands a "dark" component, usually associated to a scalar field that dominated the 
dynamics and kinematics in the very early Universe.  These three dark components are independent from each other since, at least for historical reasons,
they were invented to solve different cosmological/astrophysical problems.   However, there are many models that aim to unify the dark components of the 
Universe, all three or in pairs. In the present work, we have updated and tested further two different unification approaches: the dark fluid and an $F(X)$ scalar field. 

The dark fluid is constructed to have exactly the same properties as both dark matter and dark energy. 
Thus, with a single fluid we successfully achieve to reproduce exactly the same dynamics of the $\Lambda$CDM standard model of cosmology. Given the dark degeneracy, 
there is no way to distinguish, through dynamical or kinematical computations, between the dark fluid and the dark components of the $\Lambda$CDM model.  Our  
proposal implies that dark matter and dark energy do not separately exist, but they constitute a single fluid. We have not analyzed what the possible real candidates for the dark fluid are, but speculating,  
it could be a cold dark matter particle with an intrinsic (remanent?) small pressure.  The dark fluid could also be a collection of barotropic fluids that comply with an effective null sound 
speed propagation, and again with an associated small pressure. On the other hand, if we add baryonic interactions to the dark fluid, one breaks the degeneracy with $\Lambda$CDM and 
one can compute the constraints imposed from recent cosmological probes, as we have done in this work.      

A triple unification of dark matter, dark energy, and inflation can be carried out by using a particular $F(X)$ Lagrangian and a typical potential term. The first term is necessary to emulate
the dark matter behavior in the cosmological evolution.  Although we used a very particular model for this aim, we remarked that a cosmological dark matter behavior 
is achieved with a $F(X)$ scalar field that has a minimum (with $X_{0} \neq 0$) and it stays near to it, and thus it complies with the conditions given by equations (\ref{dm1}) and  (\ref{dm2}), that is, being 
a "fluid" with small pressure and speed of sound.  The inflationary part is achieved through any standard potential, associated to the scalar field, but some corrections apply stemming 
from the non-standard kinetic part. In our particular case, this leads to the prediction of a more reddish than measured spectrum of the primordial seeds for perturbations. Ways to change this last 
conclusion are to be worked out yet.   Finally, dark energy is put by hand, but not exactly as in the  $\Lambda$CDM model,  here a different magnitude for the "cosmological constant" is needed 
to be subtracted from a remaining constant of the dark matter part of the Lagrangian. Yet, both constants have to cancel out in a fine tuning way to yield the correct dark energy dynamics.

\begin{theacknowledgments}
JDS gratefully acknowledges financial support from the Instituto Nacional de Investigaciones Nucleares (ININ).
\end{theacknowledgments}

\bibliographystyle{aipproc}   

\bibliography{bibliography}{}

\end{document}